\documentclass[12pt]{article}
\usepackage[dvips]{graphicx}
\usepackage{floatflt}
\usepackage{color}
\usepackage{colortbl}

\newcommand{\be}{\begin{equation}}
\newcommand{\ee}{\end{equation}}
\newcommand{\bean}{\begin{eqnarray}}
\newcommand{\eean}{\end{eqnarray}}
\newcommand{\bea}{\begin{eqnarray*}}
\newcommand{\eea}{\end{eqnarray*}}

\newcommand{\bc}{\begin{center}}
\newcommand{\ec}{\end{center}}

\suppressfloats[!]

\begin{document}
   

\title{A REMARK ON THE METHOD OF ELECTRON BEAM ENERGY
           MEASUREMENT USING LASER LIGHT RESONANCE ABSORPTION.}


\author{ N.B.~Skachkov,\\
Joint Institute for Nuclear Research,
Dubna, 141980, Russia\\
E-mail:skachkov@jinr.ru, skachkov@fnal.gov}

\maketitle


%
%
%


\begin{abstract}
\noindent 


     The problem of measuring of the electron beam energy by help
     of the laser light interaction with the electrons is discussed.
     It is shown that the orthogonal  orientation of the laser
     beam with respect to the electron one, proposed in the 
     present Note, may  allow  to perform this  measurement in 
     accordance  with the  physical nature of a formation of an  
     electron quantum levels in a magnetic field. In result,
     the final formula, that expresses the beam energy through
     the strength of a magnetic field and the energy of the laser
     photon, gets a transparent physical meaning and do contain
     a less number of parameters (what may lead to an increase 
     of the precision of the measurement). Some other  sequences 
     from  this proposal, like the change of the  geometry of
     the  experimental set-up  and the necessity of a new
     additional  detector to register  the products of the 
     Compton scattering for monitoring  of the beam energy 
     measurements,  are discussed also.

\end{abstract}

\section{Introduction.}
\label{intro}


~~~~~~An  application of the resonance absorption (RA) of the 
   light by electron \cite{Ryaz} -\cite{BarMel}
   for the  beam energy  measurements
   was discussed in a number  of talks given  at different  
   conferences \cite{Melik/Protv},  \cite{BarMel}.
 
   The idea of this method is based on the theoretical formula
   which was obtained for the first time in the framework
   of the non-relativistic  quantum mechanics (see \cite{Land}) 
   and latter on was generalized for  the relativistic 
   case basing upon the Dirac and the Klien-Gordon 
   equations (see, for instance \cite{Sokolov-Ternov}).
   These formulae give the energy eigenvalues and the wave 
   functions of the particle which  moves in the homogeneous
   magnetic field. The analogous exact solutions were found 
   also for a more complicated cases when a particle
   moves in the combined field, composed of an electromagnetic
   and a homogeneous magnetic fields (Volkov solution, see also 
   \cite{Redm}, \cite{Berg}). Recently this study was continued  
   by finding out the exact solution of the Dirac equation for a
   case of the superposition of a homogeneous magnetic field and
   a circularly polarized electromagnetic wave \cite{R.A.Mel}
   with an  account of the electron anomalous magnetic moment.
   Its was shown that taking account of the electron anomalous
   magnetic moment removes the spin degeneracy of the energy 
   levels.

   According to the results of all these papers the particle
   which moves in a constant homogeneous magnetic field 
   should have the quantized values of the energy, connected
   with its motion  in the plane perpendicular to its velocity
   vector. In  \cite{Ryaz} and latter on in \cite{R.A.Mel} it 
   was proposed  to use the transition 
   from the lower levels of quantized transverse energy to
   the higher ones, caused by the laser photons absorption, 
   for getting  out the information about the initial electron 
   beam energy (let us mention that in these papers the
   process of laser photon interaction with the beam electron  
   was described within the Classical Electrodynamics 
   framework).

   In the present Note some new points of  views on this 
   measurement are proposed. First, the suggestion of a new 
   orthogonal disposition of the laser beam with respect to
   the electron one is presented. Such a geometry of  the 
   measurement is  different  to the previously considered 
   case, see  \cite{Ryaz} and  \cite{BarMel}, where the laser
   beam was supposed to be
   injected at a small angle to the electron beam direction
   (taken as z-axis).  New  proposal  is based on the arguments 
   which  follow from the physical origin of the quantized  levels 
   of the  electron energy in a magnetic field and would be 
   discussed below in Section 2. The second point is connected 
   with  the usage of  Quantum Electrodynamics (QED ) approach 
   for  describing the laser photon interaction with the  beam 
   electron (see Section 3). 
   The importance of the  measurement of the Compton
   scattering (as the process that may happen when the
   experimental parameters would not meet the conditions
   necessary for RA) for  monitoring of  the beam energy 
   measurement is discussed in the Sections 4 and 5.

\section{The energy levels and the wave function of the
        electron in the static magnetic field.}


   Here (having in mind  the aim of a further application 
   for the beam energy measurements at ILC, \footnote{it
   seems to be very useful to perform first the analogous 
   experiments at lower energies.}   i.e. at the
   ultra-relativistic energies of electron) the solution of 
   the Dirac equation would be used ( taken from the widely 
   used text book \cite{Akhi}), derived  for a case of the
   electron motion in the homogeneous  magnetic field 
   ( $\vec{B}=rot\vec{A}$),
   and with the choice of the Landau gauge
   for the 4-vector of the electromagnetic potential$ A_{\mu}$:
   $A_{0}= A_{x}= A_{z}=0$,  $A_{y}=B_{0}x$. 
   Thus, $\vec{A}=B_{0}x\vec{e_{y}}$, where  $\vec{e_{y}}$ is
   the unit vector along the y-axis,
   and $\vec{B}=(0,0,B_{z})$.

   The  energy of the electron $p^{0} $, which is the fourth component
   of the electron 4-momentum vector  
   $P_{\mu}=(p^{0}, p^{x}, p^{y}, p^{z} )$,
   is defined in this case  by formula ( \cite{Akhi} ):

\begin{equation}
  {(p^{0})^2 \equiv  E^{2}_{\lambda}(n,p_{z} ) =
    E^{2}_{z} + E^{2}_{T,\lambda}(n)},
\end{equation}
    where the first term 

\begin{equation}
{E^{2}_{z}
 = m^{2}_{e}c^{4} +{p^{2}_{z}}c^{2}}
\end{equation}
   is the square of the relativistic energy of a free electron
      that moves in a beam (taken as the z-axis).
The second term 
\begin{equation}
{E^{2}_{T,\lambda}(n) =
      h(\frac{eB_{0}}{m_{e}})(m_{e}c^{2})(2n +1 +2\lambda)} 
\end{equation}
   is the square of the relativistic energy of the  electron
   transversal motion, which depends on the strength of the
   magnetic field $B_{0}$ as well as on the spin projection
   ``$\lambda$'' on  z-axis.  Hence, the transversal energy 
   of the electron has the quantized  values, numerated by
   the  main quantum number ``n'' (n=0,1,2,3...).
   Also $m_{e}$ is the electron mass, 
   e is its charge and $h$ is the Plank constant, while  
   $\frac{eB_{0}}{m_{e}}=\omega_{c}$ is the cyclotron
   frequency.

     In what follows we shall use the expression for the
   difference of  $E^{2}_{T,\lambda}(n)$  for the two
   neighboring values of energy levels with $n+1$ and
   $n$ quantum numbers which stems from the formula (3):

\begin{equation}
 {\Delta{E^{2}_{T,\lambda }}= 
    E^{2}_{T,\lambda}(n+1) - E^{2}_{T,\lambda}(n)  = 
    2h(\frac{eB_{0}}{m_{e}})(m_{e}c^{2}) =
     2h\omega_{c}(m_{e}c^{2}). }
\end{equation}

    Now, let us  note that in a case of
   the spin projection $\lambda=-\frac{1}{2}$
   the ground state (i.e., n=0)
   would have, as it  follows from (3), the value of 
   the transversal energy been equal to zero:
   \footnote{ The next one level also having  n=0, but 
   with the spin projection $\lambda=+\frac{1}{2}$,
   would have the energy
${E^{2}_{T,\lambda=+\frac{1}{2}}(0) =
        2h\omega_{c}(m_{e}c^{2}) }$ ,
   in accordance with formulae (3), (4).}
\begin{equation}
{E^{2}_{T,\lambda=-\frac{1}{2}}(0) =0.} 
\end{equation}

   Here from comes an  important fact (which would be used 
   in what follows) that for the ground 
   state with n=0 and the spin
   projection $\lambda=-\frac{1}{2}$ the expression for the 
   \textit{total energy  of the electron in magnetic field}
   do coincides  with \textit{ the energy of  a  free electron}:
\begin{equation}
  {E_{\lambda=-\frac{1}{2}}(n=0,p_{z})  = E_{z}}=
  \sqrt{ m^{2}_{e}c^{4} +{p^{2}_{z}}c^{2}}.
\end{equation}


     In a complete correspondence with the formula (1), that
   demonstrates  the independent entrance of  $E^{2}_{z}$ and
   $E^{2}_{T}$  terms into the expression of the square of the 
   total energy  $E^{2}_{n,\lambda}$, {\footnote{ there is no 
   interference term !} the wave functions of the Dirac
   equation,  found as the exact solutions for the mentioned 
   above different  combinations of  external magnetic 
   and electric fields \cite{Ryaz}- \cite{Akhi}, also do
   factorize in two parts.
 
   One part (an oscillator like solution) describes the motion 
   in the plane transverse to the beam and along the x -axis
   (because the only  nonzero component  $A_{y}=B_{0}x$ of the
   4-vector of the  electromagnetic potential $A_{\mu}$ is 
   defined by the x-coordinate).
   The another one, which  has an exponential  form, 
   is  connected with the free  motions along the y- and
   z-axis (see  \cite{Land} -   \cite{Akhi} ).
   For example,  the solution, that corresponds to the formula 
   (1),   has  a form ($\vec{r}=(x,y,z)$) \cite{Akhi}  :
\begin{equation}
  {\psi(\vec{r})= e^{i(p_{y}y+p_{z}z)}f(x),}
\end{equation} 
   where
\begin{equation}
   {f(x)=Ce^{-{\xi^2}/{2}}}H_{n}(\xi); 
     \xi=\sqrt{eB_{0}}(x-p_{y}/eB_{0}).
\end{equation}  
   Here C is the normalization constant and $H_{n}(\xi)$ is
   an Hermite polynomial which appear is a typical component
   of many solutions in a case of an oscillator type potentials.
   The exponents in $\psi(\vec{r})$ are the plane wave
   functions that do represent the typical solutions which 
   describe a free motion of particles. 

    So, after we have discussed  what type of motion is
   described by different components of the wave function, 
   let as recall that in quantum mechanics the square of 
   the wave function gives the value of the  probability of
   finding out the particle at some point $\vec{r}$.
   Let us also add  that from the  statistics it is  known 
   that the probability of 
   the process, that consists of the set of independent 
   events, is defined as the product of the probabilities
   of independent events. As it is seen from the formula
   (7) this is just our case.

     In this way  the apparatus of quantum mechanics 
   demonstrates that the motion of the electron in the 
   homogeneous  magnetic field does consists in reality of 
   the  sum of two independent motions. One motion, which
   is  performed along the z-axis, is a free longitudinal
   motion and it is  characterized by the momentum $p^{z}$.
   Another one is performed  in the x-y plane, i.e.
   perpendicular to the beam  axis, and does consists of 
   the oscillator motion along the x-axis (this is a
   preferable choice of the coordinate system in the most
   of text books) and of a free motion 
   along the y-axis with the momentum $p^{y}$.
 
   After this discussion of the physical meaning  of the
   formulae, used to describe the features of the quantized 
   energy  levels of the electron which moves in the
   homogeneous magnetic field, it is a time to consider a
   question of describing  the interaction of the laser 
   light with the beam electrons.

\section{The laser photon interaction with the electron.} 
      
     It is well known that in QED the lowest order (in e) 
   amplitudes, i.e. the amplitudes for  the  $1 \to 2$  
   processes (like, for
   instance, a free electron  transition into  a free 
   electron  and a photon, i.e. ${e \to \gamma +e} $ process,
   as well as a photon transition into the electron
   positron  pair ${\gamma \to e^{+} + e^{- } }$ ) 
   are  forbidden due to  the relativistic kinematics  
   (see, for example,  \cite{Akhi}-
   \cite{NNBog}). To this reason in QED the lowest order 
   amplitude does corresponds to the next to leading
   (i.e. $e^{2}$ order) diagrams that describe a  
   $2 \to 2$ transitions like
   ${\gamma + e \to \gamma + e}$ (Compton scattering). 
   This result is in agreement with the common sense statement
   that the electron, been forced to change his speed 
   (accelerated/deaccelerated) by some external influence, 
   should radiate the energy. In QED it means an emission 
   of photons.
   
    In a correspondence with this statement
   the 4-momentum conservation law  for
   the Compton process of the photon (with the 4-momentum 
   $k_{1}$) scattering off the electron (with the 
   4- momentum  $p_{1}$) looks, in a general case, like
   ($k_{2}$ and  $p_{2}$ 
   are the final state momenta of the  photon and the
   electron):


\begin{equation}
{p_{1}+k_{1} = p_{2}+k_{2}. }
\end{equation}  
  If we shall pass to the  components of the 4-momentum:
  $k^{\mu} = (k^{0},\vec{k})$; 
  $\vec{k}=(\vec{k_{T}},k^{z} )$, then the equation
  (9)  may be presented  as a set of the independent
  conservation laws for each  component separately:\\

  for the energy, 
\begin{equation}
{p_{1}^{0}+k_{1}^{0} = p_{2}^{0}+k_{2}^{0} },
\end{equation}    

  for the z-projection:
\begin{equation}
{p_{1}^{z}+k_{1}^{z} = p_{2}^{z}+k_{2}^{z} },
\end{equation}    

and for the transversal one

\begin{equation}
{\vec{p_{1}^{T}}+\vec{k_{1}^{T}}=\vec{p_{2}^{T}}+\vec{k_{2}^{T}}}.
\end{equation}    

   Thus, if the laser photon would have some small angle 
  $\theta$ to the  electron beam,  as it was considered in  
  \cite{Ryaz} and  \cite{BarMel}  (in such a case 
  the $k_{1}^{z}$ component of the photon  momentum would 
  dominate  over his transverse one $k_{1}^{T}$) ,
  then the z-component of the  electron would change, in
  general case, from $p_{1}^{z}$ to some $p_{2}^{z}$ due to 
  the conservation law of the  z-component of 4- momentum.
 
    Independently from the z-component,  the transversal
  component of electron may also change after the  interaction. 
  There  are two possible cases, which  may be realized in the 
  framework of QED.

    The first one, most simple, may be realized  when the 
  $\vec{k_{1}^{T}}$  contribution to the electron transversal
  motion would not  match the difference (4) of two quantized
  electron energy  levels in a magnetic field. In this case, 
  according to QED a pure Compton  
  scattering would take a place.

   The second one
  may happen when the value of  $|\vec{k_{1}^{T}}|$  should
  match the equation(4).
  In this case, keeping in mind what was written before about
  the independent conservation of z- and transverse components 
  of the  total 4-momentum of electron-photon system, we
  shall have the final state consisting of :

   . the photon, radiated due to changing of the electron
      momentum z-component (and  \\
~~~~~ in accordance with (11)), 
  
   . and the electron,  which $|\vec{p_{2}^{T}}|$ should be
     defined in this case  by formula (3) for the \\ 
~~~~~ (n+1)-th , or even higher level, 
     if it was at the  n-th orbit in the initial state.

  Hence,  in both cases, 
  \textit{ even if the resonance condition would
  be fulfilled   and the transverse component $k_{1}^{T}$ would
  exactly coincide with the energy necessary for moving 
  from the lower orbit to the higher one}
    (i.e. the equality
  $\vec{p_{2}^{T}}=\vec{p_{1}^{T}}+\vec{k_{1}^{T}}$ 
  would be fulfilled for the
 \textit{independent(!) motion in the transverse plane}), 
 \textit{the changing of the
  independent z-component of the electron momentum  would
  lead to the  radiation from the electron}. In QED it is 
  an emission of the photon via the $e^{2}$ order Compton 
  process.
       \footnote {which would have the energy and angular
        distributions similar to those that appear in a
        case of laser back scattered photons, described 
        in  \cite{TESL_TDR_gg}.
  It should 
  be mentioned that the estimates of the laser photon
  energy, or, of the wave length, appropriate for the RA
  effect at TESLA  energies  \cite{BarMel}, give the value 
  about of $\lambda_{las}=10{\mu}m$. This number is quite
  comparative with the  wave length of the laser 
  $\lambda_{las}=1 {\mu}m$,  planned to produce the
  backscattered laser photon 
  beam for photon-photon collisions
\cite{TESL_TDR_gg}.} 
 
   In other words, according to QED, in a case, when the
  laser  beam would have a small angle $\theta$ with
  respect to the  electron beam,  the Compton process 
  (with or  without moving of an electron from the lower
  orbit to  a higher one)
  would take place  and  the  emission of the
  photon would happen. In such a case the whole
  situation  becomes more complicated for the theoretical
  description. \footnote{there are
  no complete calculations found in the literature that may
  describe this general case  up to the end.} 

  A  possible simple way to avoid these problems and to use 
  the theoretical  prediction about the existence 
  of the quantized energy levels (of the transverse motion
  of the  particle in the magnetic field)  for the electron
  beam measurement  is given below. It is based on a special
  choice of  the angle  between the laser and the electron 
  beams, mentioned already in the Introduction.

\section{What laser photon  beam orientation meets better 
    the  physics of electron motion in magnetic field?}

  The physical picture of the electron motion in the static 
  magnetic field as been a  sum of two independent motions 
  (what was discussed in the previous Sections) as well as the 
  4-momentum conservation equations written above, do prompt 
  (if not to say, dictate) an easy way to prepare the initial
  state to use the possibilities connected with existing of
  the  quantized energy levels of a particle in the magnetic 
  field. This possibility is based on the proper choice of 
  the relative  geometrical orientation of the laser and 
  the electron  beams. 
    
    Indeed, if we shall choose
  the laser beam to be orthogonal to the electron one,
  i.e. shall consider a case when  $k_{1}^{z}=0$ in
  equation (11),  then the collision 
  of the laser photon with the electron would not effect the
  z-component of the electron momentum at all.

    If also, in  addition to the orthogonal angle orientation,
  we shall adjust the laser
  photons energy (and the magnetic field strength  $B_{0}$)
  \footnote{ better to say to adjust the choice
  of the laser to get the most suitable wave length and then 
  to tune the value of the magnetic field strength $B_{0}$. 
  We shall return to this question in the next Section.}
  to be equal to the difference of the two lowest quantized
  levels of energy, connected with the transversal motion of
  the electron in a magnetic field and defined by formula 
  (4),  then the laser photon would be absorbed.
      \footnote{ let us mention  that in the framework of 
       QED, the process of photon absorption goes on
       practically immediately.}
  In this case the whole initial  photon  energy, been  
  prepared  as of the transversal  nature only, would
  transform into the 
  energy of a its  circular  motion, but  on a higher then
  previous one  orbit in the plane transverse 
  to electron beam.  Therefore, in a case of the proposed 
  orthogonal orientation and the properly tuned value of the
  magnetic field strength $B_{0}$ in a way to have the 
  difference of two energy levels of the electron in 
  accordance with the energy of the laser photons 
  \textit{there would  be no any phase space left for the 
  production of an additional photon in the final state}!

   In this special case of $k_{1}^{z}=0$ and when also the
  laser energy would fit 
  the difference of two $E_{T}$ for some n and n+1
  levels, the formula (11) shall take a very  simple form:

\begin{equation}
{p_{1}^{z} = p_{2}^{z} }.
\end{equation}    
    (This equation may be treated also as one of the 
  conditions of RA effect.)  

   The energy conservation law in the such a  case of 
 $\theta_{laser}=90^{0}$  should look  like

\begin{equation}
{p_{1}^{0}+k_{1}^{0} = p_{2}^{0} },
\end{equation}
  where, due to our choice of the laser photon momentum
  4-vector in a form 
  $k^{\mu}_{1} = (k^{0}_{1},\vec{k_{1}})$;\\ 
  $\vec{k_{1}}=(\vec{k_{1T}}, 0 )$, 
  the final state electron transverse
  momentum  $p_{2}^{T}$ would be defined by the 
  the equation 
\begin{equation}
 {\vec{p_{1}^{T}} + \vec{k_{1}^{T}} = \vec{p_{2}^{T}} },
\end{equation}    
  as the sum of the initial electron  transverse momentum 
  $\vec{p_{1}^{T}}$ and the laser photon transverse vector
  $\vec{k_{1}^{T}}$  which modulus is equal to the energy 
  of the  absorbed photon
  \footnote{i.e., $k^{0}_{1}=|\vec{k_{1T}} |$ due to the relation
  $k^{0}_{1}=|\vec{k_{1}}|$, which is valid for the 
  massless photon.}.

  The 4-component equation (9)
  would take in such a  case the following form:
\begin{equation}
{p_{1}+k_{1} = p_{2} }.
\end{equation}  

  Note that for the considered  here process the right
  hand side of equation (14) may be represented like
\begin{equation}  
p_{2}^{0}=\sqrt{ E^{2}_{\lambda}(n+1,p^{z}_{2}} )
\end{equation}
  (where the value $ E^{2}_{\lambda}(n+1,p^{z}_{2})$ 
  is  defined  by  formula (1))  if  the beam electron 
  initially  was  ``on the n-th orbit''.

   Taking the square of equation (14) and with an account 
  of formula (1)  one comes
   to the equation for a non spin-flip case


\begin{equation}  
 {E^{2}_{\lambda}(n,p_{1}^{z} ) +
   2k^{0}_{1}E_{\lambda}(n,p^{z}_{1} ) +
      (k^{0}_{1})^{2}=
         E^{2}_{\lambda}(n+1,p^{z}_{2} )},
\end{equation}  
    which can be rewritten (in what follows we shall  use  
    the notation  $k^{0}_{1}=E_{las}^{\gamma}$) like
\begin{equation}  
 {2E_{las}^{\gamma}E_{\lambda}(n,p_{1}^{z} ) +
     (E_{las}^{\gamma})^2=
        E^{2}_{\lambda}(n+1,p_{2}^{z})-
            E^{2}_{\lambda}(n,p_{1}^{z}).}
\end{equation}
   This is a general form of equation (14) for any 
   values of $p_{1}^{z}$ and  $p_{2}^{z}$. 
    
  Now, if we shall take into account the $p^{z}$ conservation 
  law in a form (13), that corresponds only to the
  discussed above laser to beam orthogonal orientation,
  then the right hand side of (19) would contain  the
  expressions  of the total energies of electrons in 
  the  magnetic field for a case of equal values 
  of  $p^{z}$ components ($p_{1}^{z} = p_{2}^{z}$):

\begin{equation}  
 {2E_{las}^{\gamma}E_{\lambda}(n,p_{1}^{z} ) +
     (E_{las}^{\gamma})^2=
        E^{2}_{\lambda}(n+1,p_{1}^{z})-
            E^{2}_{\lambda}(n,p_{1}^{z}).}
\end{equation}
   From formula (1)-(3) it is clear that the longitudinal
   contributions $E^{2}_{z}$ would cancel in the right hand
   side of (20)
    \footnote{ here the advantage of the orthogonal beams
     orientations, accumulated in equation (13), come
     into a play! }  
   and only  the difference of the transverse
   energies  $E^{2}_{T,\lambda}(n)$ would remain  there:
\begin{equation}  
 { E^{2}_{\lambda}(n+1,p_{1}^{z})-
            E^{2}_{\lambda}(n,p_{1}^{z})=
      E^{2}_{T,\lambda}(n+1) - E^{2}_{T,\lambda}(n). }
\end{equation}


 The expression for this
   difference was already given by formula (4). 
 So formula
   (20) could be presented in a form
\begin{equation}  
 {2E_{las}^{\gamma}E_{\lambda}(n,p_{1}^{z} ) +
     (E_{las}^{\gamma})^2=
       \Delta{E^{2}_{T,\lambda }}, }
\end{equation}
   which  does not contain any more the dependence
   on the beam energy in the right hand side of the
   equation. In such a case it is convenient to
   present  (22) as an equation which includes the 
   dependence on  the electron beam energy  (see (1))
   only in the left hand side: 
\begin{equation}  
 {E_{\lambda}(n,p_{1}^{z})=
       \frac{\Delta{E^{2}_{T,\lambda}}-
           (E_{las}^{\gamma})^2}{{2E_{las}^{\gamma}}}.}
\end{equation}

    Let us  consider now  the case of the non spin-flip
   transition from  the ground state n=0 to the state
   with  n=1 (we shall choose a case
   when both states have  
   $\lambda=-\frac{1}{2}$),  then the  formula (23) 
   (written as
   ${ E_{\lambda=-\frac{1}{2}}(n=0,p_{z})  = E_{z}}=
  \sqrt{ m^{2}_{e}c^{4} +{p^{2}_{z}}c^{2}}= E_{beam})$
   gives,  due to (6), the following expression for the
   beam  energy $E_{beam}$ 
\begin{equation}  
  { E_{beam}=
  \frac{\Delta{E^{2}_{T,\lambda=-\frac{1}{2}}}-
        (E_{las}^{\gamma})^2}{2E_{las}^{\gamma}}.}
\end{equation}
   Keeping in mind that according to formula (4)
   for the non spin-flip case,  i.e. when the value of
   $\lambda$ does not change in the process of the beam
   electron interaction with the the laser photon, we have 
    $\Delta{E^{2}_{T,\lambda }}= 
        2h\omega_{c}(m_{e}c^{2})= 
            2h(eB_{0}/m_{e})(m_{e}c^{2})$, 
   we come to a very simple final formula 
\begin{equation}  
  { E_{beam}=
  \frac{2h(\frac{eB_{0}}{m_{e}})(m_{e}c^{2}) -
         (E_{las}^{\gamma})^2}{2E_{las}^{\gamma}}.}
\end{equation}
   The last equation does express the value of the  beam
   energy through those
   values of the  magnetic field  strength $B_{0}$ and of
   the laser photon energy $E_{las}^{\gamma}$,  \textit{ which 
   were tuned to provide the conditions for the laser photon 
   energy to be equal to the  difference of two neighboring 
   quantum  energy levels in a case of the orthogonal relative 
   orientation  of the laser to the electron beams.}

   It is natural that \textit{ the error of  the measurement of 
   the electron beam  energy $\Delta{E_{beam}}$  should be
   defined in this case by  two errors only :\\ \\
~~~~  .  the  error of the  magnetic field 
   strength $\Delta{B_{0}}$ measurement, and by\\ \\
~~~~  .  the error of the measurement of the laser photon energy
   $\Delta{E_{las}^{\gamma}}$ . }\\

    To summarize shortly it is possible to say that in
   a case when the following conditions would be
   fulfilled: \\

   \textit{*~ the angle between the laser  and the 
    electron beams  would be chosen to be  $\theta=90^{0}$,
    (i.e. $cos\theta=0$)},  

   \textit{** the  magnetic field  strength $B_{0}$ and
   the laser  photon energy $E_{las}^{\gamma}$ 
   would be  tuned  to  provide the conditions for the 
   laser energy to be equal to the  difference of two
   neighboring quantum energy levels of the electron 
   in a homogeneous magnetic field },\\

   then the  photon may be absorbed  by the quantum system
  of the quantized  electron levels in a  homogeneous
  magnetic  field and \\

  \textit{ the laser photon collision with the beam electron
  may lead only
  to a change of the orbit number of the transverse motion 
  of the electron. No changes of the longitudinal (i.e., 
  of z-)  component of the electron momentum would happen.
  It means that no any additional radiation (like 
  that one which take place  in the Compton scattering 
  process) would have sufficient phase space to appear}.


\section{How it may look from the experimentalist point of view. }



     There are some points have to be discussed  in a
   connection with a possible planning of the measurements
   if they should be done in the context 
   of the described  above picture of the laser photon  
   interaction with the beam electron.


    The obvious question needed to  be asked for such a
   planning would be about the choice of the laser energy
   (i.e.,  the wave length). The answer can be easily found
   from the solution of  the quadratic equation (25) for
   the $E_{las}^{\gamma}$. In  a case of n=0 and
   $\lambda=-\frac{1}{2}$  we come to the formula
   \footnote {  in a case when 
   $\Delta{E^{2}_{T,\lambda}}/{E_{beam}^{2}}< 10^{-2}$
   this formulae simplifies and turns into a very easy
   relation 
   $E_{las}^{\gamma}=\Delta{E^{2}_{T,\lambda}}/2E_{beam}$.  
   It is  rather curious to note that
   its inversion gives  a more strait forward  (but a bit
   approximate) expression for the electron  beam energy
   $E_{beam}=\Delta{E^{2}_{T,\lambda}}/2E_{las}^{\gamma}$,
   which, by the way,  may be obtained  from the
   equation(23)  if one   shall neglect
   the term $(E_{las}^{\gamma})^2$ in its numerator.The
   similar approximation was used in  \cite{Ryaz},
   and  \cite{BarMel} for a case of $cos\theta \not= 0$.}

\begin{equation}
  {E_{las}^{\gamma}=E_{beam}(\sqrt{ 1+      
 \frac{2h(\frac{eB_{0}}{m_{e}})(m_{e}c^{2})}{E_{beam}^{2}}} -1)
    =E_{beam}(\sqrt{ 1+      
      \frac{\Delta{E^{2}_{T,\lambda}}}{E_{beam}^{2}} } -1)},
\end{equation}
   which  expresses  $E_{las}^{\gamma}$ through the value
   of the beam energy $E_{beam}$ and the strength of the
   magnetic field $B_{0}$. So, with the help of this formula
   one can find out (for a chosen value of  $E_{beam}$ and
   an expected  value of  $B_{0}$) what kind of a laser does
   have the  wave length that corresponds to the energy 
   defined  by formula (27).


    The next important sequence for planning of the
   experiment is  the need to pass to  a new geometry for
   the experimental  set-up for  such  measuring of 
   dropping of  the laser light  intensity due to the
   photon absorption.
   It is evident that the   detector (D), proposed 
   to  measure the decrease of the  laser light
   intensity  after  its interaction with the 
   electron beam  \cite{BarMel}, and the laser  should be 
   placed on the  same line, perpendicular to the electron
   beam . Such a  geometrical disposition of the experimental
   setup looks very  different to what  was discussed  in
   \cite{Ryaz}- \cite{BarMel}, where it was
   proposed that the  line which connects the laser and the
   detector (D) would have a small angle $\theta$ to the 
   electron  beam line.
     
     This changing of the position of the detector D may be 
   important  from the experimental point of view. Really, 
   in principle,  there is  a non zero probability that the 
   electron, which was forced by the laser light to  move
   to a higher orbit, would emit a photon and then
   return  after some time  back onto
   the ground state orbit. Here we
   have a direct analogy with the radiative transition from
   higher to lower orbits in atoms. Fortunately, in the case, 
   proposed in the present Note, the  electron, been moved
   onto  any the higher orbit, would keep, as it was
   discussed in the previous  Sections, 
   the same   value of the initial longitudinal momentum 
   $p_{1}^{z}$ (as it was discussed in the previous 
   Sections), i.e., it  shall continue to move with the same
   speed,  practically equal to  the speed of light. 
   Hence, if the photon  emission would  happen at some
   moment, defined by the  lifetime of the ``n+1-th''
   excitation level, nevertheless, this photon 
   have no chances to hit the detector D , 
   because during  this even small time the
   electron, that was pushed to the ``n+1-th''
   orbit,  would fly already away from the position of the  
   ``laser-detector D'' perpendicular line (to the electron 
   beam line).
   To this reason the proposed here perpendicular orientation
   of the laser to electron beam  is a  preferable one 
   as it is more safe from the view point of the possibility 
   that the secondary radiative 
   photon emission may reach the detector D.
 
    Up to now we were discussing the aspects connected with
  the case of the  resonance absorption within the situation
  of  a new orthogonal
  disposition of both beams. Now let us turn to the situation 
  when the energy of the laser photon would not fit to the
  the value of energy splitting  between the main n=0 and
  one of the next to the ground states levels. 
     
    As it was already discussed in the  previous Sections,
  in a case when the resonance condition (25) would not be
  fulfilled (most probably,  due to some variation of the 
   electron beam energy) then according to
  QED the Compton process of ${\gamma + e \to \gamma + e}$ 
  scattering would  take place. 

    The appearance of the Compton scattering products,
  i.e. of the final state photon and electron (with
  $p_{2}^{z} \not= p_{1}^{z}$) may
  be used as the signal of a dropout from the resonance
  condition (25). Therefore, the addition of experimental 
  equipment, supposed for beam energy measurement, by a new
  detector which may identify the electrons and photons,
  produced in the Compton scattering, may be very helpful.  
  It may be an electromagnetic  calorimeter with the 
  appropriate granulation  for the registration of the 
  photons and for defining of their position,  
  accompanied  by some electron tracker.\footnote{Really, 
  if there would be only one detector D for measuring of
  the decrease of the laser light, caused by the resonance  
  absorption, then we  would  not have a full information
  about the origin of such a variation, because  it may 
  happen  also, for instance, due to the variation
  of the  beam density.} 

    It should be mentioned also that  if we shall suppose
   that the effects of the electron beam energy  variation
   would not be so large, then  it would  not be so difficult
   to  calculate the angle and  energy distributions of
   the  scattered photons and electrons which  would be 
   characteristic for the such possible variations of 
   the electron beam energy in both directions from the 
   position of the resonance value.

   Thus, it may happen so that the control over the Compton 
   process (by a comparison of the observed experimental
   distributions of  the photon and the electron 
   parameters with those predicted by QED) would turn 
   out to be  a useful guide for adjusting the magnetic
   field  strength $B_{0}$ during the beam energy measurement.
   In such a case the measuring of the  characteristics of 
   the Compton scattering within the region of parameters,
   close to the resonance position, may be used for the on-line  
   tuning of the magnetic field strength to those values 
   which would be an adequate to keep the resonance
   condition. The detailed  study  of this possibility would 
   be a subject of the following publications.

%
\section{Summary.}
%

     It is shown that the orthogonal orientation  of the
   laser photon beam with respect to the electron beam
   allows to bring the idea of the use of the theoretical 
   prediction about  the existence of electron quantum levels 
   (in the presence of the homogeneous magnetic field) in the
   correspondence with the physical origin of these resonance
   levels.

    In result, the formulae that describe a process of a
   possible  absorption of the laser light by the transverse
   degrees of electron motion in the constant magnetic
   field become much simpler because in this case the
   absorption process does not interfere with the 
   longitudinal motion  along the beam direction.
   Due  to this no additional radiation would appear in
   a case when the laser photon energy would fit the value of 
   the energy  splitting of two electron levels in  the
   magnetic  field (i.e. in a case of fulfilling the
   resonance condition). In result, the physical picture
   of the process of measuring the electron beam energy
   by measuring the drop of the laser light intensity
   in the absorption process becomes more transparent.
   This new relative orientation of two beams means also  
   passing to another geometry for experimental tools set-up.

   In the opposite  case when the energy of the laser photon 
   would  not be  equal to the difference of two electron
   energy levels (in magnetic field) the Compton scattering
   process would take place. The characteristics of
   the scattered electron and photon (angles and energies)
   may be measured (it may require to enlarge the experimental
   equipment by the additional
   electromagnetic calorimeter and the tracker) and then 
   used  for an on-line monitoring of the process of tuning 
   the magnetic field to keep the resonance photon absorption
   condition (25) for the electron beam energy measurement.

      It is a pleasure for the author to express his 
   gratitude to R.Melikian and  especially to H.J.Schreiber 
   for the useful discussions.  


\end{document}